\documentclass[final,1p,times]{elsarticle} 

\usepackage{graphicx}
\usepackage{amssymb} 
\usepackage{amsthm} 
\usepackage{lineno}
\usepackage{subfigure}

\newcommand{\glabcms}{\gamma^{\rm lab}_{\rm c.m.s.}}
\newcommand{\blabcms}{\beta^{\rm lab}_{\rm c.m.s.}}

\newcommand{\ie}{{\it i.e.}}
\newcommand{\eg}{{\it e.g.}}
\newcommand{\etal}{{\it et al.}}
\newcommand{\sqrtS}[1]{\mbox{$\sqrt{s_{#1}}$}}
\newcommand{\calL}{\cal L}

\def\pp   {$pp$}
\def\pd   {$pd$}
\def\pA   {$pA$}

\def\PbA {Pb$A$}
\def\jpsi    {\mbox{$J/\psi$}}

\def\sqrtsNN {\mbox{$\sqrt{s_{NN}}$}}
\newcommand{\ct}[1]{{Table~\ref{#1}}}

\usepackage{ifpdf}
\ifpdf
\usepackage[pdftex]{hyperref}
\else
\usepackage[hypertex]{hyperref}
\fi

\hypersetup{
  pdftitle={},%
  pdfauthor={},%
  pdfsubject={},%
  pdfkeywords={},%
  pdfstartview={},%
  bookmarksopen=true, breaklinks=true, debug=true, %
  colorlinks=true, linkcolor=red, citecolor=blue, urlcolor=blue
}

\journal{Nuclear Physics A} 

\begin{document}

\begin{frontmatter} 

\title{Ultra-relativistic heavy-ion physics with AFTER@LHC}


\author[irfu]{\small A.~Rakotozafindrabe} 
\author[torino]{\small R.~Arnaldi}
\author[slac]{\small S.J.~Brodsky} 
\author[ipno]{\small V.~Chambert}
\author[ipno]{\small J.P. Didelez}
\author[ipno]{\small B. Genolini}
\author[usc]{\small E.G.~Ferreiro}
\author[llr]{\small F.~Fleuret}
\author[ipno]{\small C.~Hadjidakis}
\author[ipno]{\small J.P.~Lansberg}
\author[ipno]{\small P.~Rosier}
\author[lpsc]{\small I.~Schienbein}
\author[torino]{\small E.~Scomparin}
\author[aarhus]{\small U.I.~Uggerh\o j}

\address[irfu]{\small IRFU/SPhN, CEA Saclay, 91191 Gif-sur-Yvette Cedex, France}
\address[torino]{\small INFN Sez. Torino, Via P. Giuria 1, I-10125, Torino, Italy}
\address[slac]{\small SLAC National Accelerator Laboratory, Theoretical Physics, Stanford U., Menlo Park, CA 94025, USA}
\address[ipno]{\small IPNO, Universit\'e Paris-Sud, CNRS/IN2P3, F-91406, Orsay, France}
\address[usc]{\small Departamento de F{\'\i}sica de Part{\'\i}culas, Universidade de Santiago de C., 15782 Santiago de C., Spain}
\address[llr]{\small Laboratoire Leprince Ringuet, \'Ecole Polytechnique, CNRS/IN2P3,  91128 Palaiseau, France}
\address[lpsc]{LPSC, Universit\'e Joseph Fourier, CNRS/IN2P3/INPG, F-38026 Grenoble, France}
\address[aarhus]{\small Department of Physics and Astronomy, University of Aarhus, Denmark}

\begin{abstract} 
We outline the opportunities for ultra-relativistic heavy-ion physics  
which are offered by a next generation and multi-purpose fixed-target experiment exploiting 
the proton and ion LHC beams extracted by a bent crystal.

\end{abstract} 

\end{frontmatter} 

\section{Introduction}

Before the advent of RHIC and the LHC, relativistic collisions of heavy ions have always been studied at
fixed-target experiments. Clearly, the possibility to use heavy-ion colliders has opened  new horizons 
in terms of the study of hard probes, which can only be produced abundantly at high energies.
Yet, one should not overlook the critical advantages of the fixed-target mode for heavy-ion physics. Briefly, these advantages are 
\begin{itemize} \itemsep-1pt
\item[-] extremely high luminosities thanks to the high density of the target, 
\item[-] the unlimited versatility of the target species which allows for in-depth studies of
yields as a function of the centrality through the $A$ dependence or the nuclear length, 
\item[-] the reduced constraints of the $P_T$ and $y$ of the studied particles due to the boost 
between the laboratory frame and the c.m.s. 
\item[-]  the possibility
for thorough and ultra precise baseline studies in proton-nucleus collisions --and even in 
proton-proton collisions-- at similar energies.
\end{itemize}
These assets are particularly striking in view of the first hard-probe studies at
the LHC where 
\begin{itemize}  \itemsep-1pt
\item[-] the request for $pA$ measurements necessarily conflicts  with those for more run time to reach
higher statistics for $AA$ measurements, 
\item[-]  the low $P_T$ region --probably the most important to understand--
is not easily accessible for hard probes such as the $J/\psi$ for the CMS and ATLAS experiments, 
\item[-] the precision of centrality-dependent observables is challenged by our determination of
the so-called centrality classes, etc.
\end{itemize}

In this context, it is well worth advertising the unprecedented possibilities offered
by a fixed-target experiment using the proton and lead LHC beams extracted by a bent 
crystal~\cite{Brodsky:2012vg}. We referred to such a project\footnote{See \url{http://after.in2p3.fr}} of a next generation and multi-purpose 
fixed-target experiment as ``AFTER", standing for ``A Fixed Target ExperiRement \symbol{64} LHC''. 

In the following, we outline the opportunities for 
ultra-relativistic heavy-ion physics, which AFTER can provide. In section 2 we present some generalities
about the experiment during the LHC Pb run. In section 3 we elaborate on the more
specific case of quarkonium production. Section 4 briefly summarises our conclusions.

\section{A fixed-target experiment with the heavy-ion LHC beam extracted by a bent crystal}

Bent-crystal beam extraction is a mature technique which offers an ideal 
way to obtain a clean and very collimated high-energy beam, without altering the performance 
of the LHC~\cite{Uggerhoj:2005xz,Uggerhoj:2005ms,LUA9}. The multi-TeV LHC beams ($E_p=7$ TeV and $E_{\rm Pb}=2.76$ TeV per nucleon) grant the 
most energetic fixed-target experiment ever performed to study \PbA\  and PbH collisions at 
$\sqrt{s_{NN}} \simeq 72\,\mathrm{GeV}$ ($\sqrt{2E_{\rm beam} m_N}$) as well as \pp, \pd\ and \pA\ collisions 
at $\sqrt{s_{NN}} \simeq 115\,\mathrm{GeV}$. As regards the heavy ion case, 
first tests on lead-beam collimation/extraction at the SPS through a 50 mm bent crystal
showed the feasibility of a large-angle deflection~\cite{Arduini:1997nb}. Recently, further tests of 
a small-angle deflection with a shorter crystal --2mm-- have also been performed successfully~\cite{Scandale:2011za}. 
In addition, a new technique 
to bend diamond crystal --thus extremely tolerant to high radiation doses-- using laser ablation 
techniques has been shown to be successful~\cite{Balling:2009zz}.

The intensity of the extracted lead beam from the LHC by putting
a bent crystal in the halo of the circulating beam can easily reach
$2 \times 10^5$ Pb/s~\cite{Uggerhoj:2005xz,Uggerhoj:2005ms}. Over a 10-hour fill, this corresponds to an extraction of about 15 \% of 
the lead ions contained in the beam ($4.1 \times 10^{10}$), which would be lost in the collimators anyway. 
The extracted beam will show a similar
bunch structure as that of the one circulating in the beam and one expects to extract on average 0.03 ions from
each bunch at each pass. No pile-up is therefore expected.
Correspondingly, the typical instantaneous luminosities achievable in 
\PbA\ mode with 1cm targets range from 7~mb$^{-1}$s$^{-1}$  for Pb to 25~mb$^{-1}$s$^{-1}$ for Be 
(\ct{tab:yieldsPb} (a)). The 
yearly luminosity for PbPb (7~nb$^{-1}$) is thus twice as large as the one expected
at RHIC~\cite{PHENIXdecadal}  in AuAu at 200 GeV and 60 times that at 62 GeV. It is also more than 10 times that expected at the LHC. 
As discussed later, despite the smaller c.m.s energy, 
the hard-probe yields expected at AFTER in PbPb are also competitive with RHIC and LHC experiments.

\begin{table}[!hbt]\footnotesize
\centering \setlength{\arrayrulewidth}{.8pt}\renewcommand{\arraystretch}{1.1}
\subtable[Instantaneous and yearly luminosities]
{
\begin{tabular}{ccccc}
\hline\hline
Target&$\rho$&$A$&$\calL$ &$\int\calL$ \\
 &(g cm$^{-3}$)&&(mb$^{-1}$ s$^{-1}$)&(nb$^{-1}$ yr$^{-1}$)\\
\hline 
10 cm liquid H & 0.068 & 1   & 80  & 80  \\
10 cm liquid D & 0.16  & 2   & 100 & 100 \\
1cm Be       & 1.85  & 9   & 25 & 25 \\
1cm Cu       & 8.96  & 64  & 17 & 17 \\
1cm W        & 19.1  & 185 & 13 & 13 \\
1cm Pb       & 11.35 & 207 & 7  & 7  \\
$d$Au {\scriptsize 
 (200 GeV)}     &   -- & --  & -- &150 \\
$d$Au {\scriptsize 
 (62 GeV)  }    &  --& --  &  -- &3.8  \\
AuAu {\scriptsize 
 (200 GeV)}     & --& --   &  -- &2.8   \\
AuAu {\scriptsize 
 (62 GeV)}      &  --&  --  & -- &0.13\\
$p$Pb {\scriptsize 
 (8.8 TeV) }     &   --&  --  &100 &100  \\ 
PbPb {\scriptsize 
 (5.5 TeV) }     &   --& --   & 0.5 &0.5   \\ 
\hline\hline 
\end{tabular}
}
\subtable[\jpsi\ and $\Upsilon$ inclusive yields]{\begin{tabular}{ccc}

\hline\hline
Target       & $N_{J/\psi}$&
$N_{\Upsilon}$
\\
  & (yr$^{-1}$)& (yr$^{-1}$)\\
\hline
10 cm liquid H         & 3.4 10$^5$   & 6.9 10$^2$ \\
10 cm liquid D       & 8.0 10$^5$   & 1.6 10$^3$ \\
1 cm Be                & 9.1 10$^5$   & 1.9 10$^3$ \\
1 cm Cu                 & 4.3 10$^6$   & 0.9 10$^3$ \\
1 cm W                  & 9.7 10$^6$   & 1.9 10$^4$ \\
1 cm Pb                & 5.7 10$^6$   & 1.1 10$^4$ \\
$d$Au {\scriptsize 
 (200 GeV)}     & 2.4 10$^6$   & 5.9 10$^3$ \\
$d$Au {\scriptsize 
 (62 GeV)  }     & 1.2 10$^4$   & 1.8 10$^1$ \\
AuAu {\scriptsize 
 (200 GeV)}      & 4.4 10$^6$   & 1.1 10$^4$ \\
AuAu {\scriptsize 
 (62 GeV)}      & 4.0 10$^4$   & 6.1 10$^1$ \\
$p$Pb {\scriptsize 
 (8.8 TeV) }      & 1.0 10$^7$   & 7.5 10$^4$ \\ 
PbPb {\scriptsize 
 (5.5 TeV) }     & 7.3 10$^6$   & 3.6 10$^4$  \\
\hline\hline
\end{tabular}}

\caption{(a) Luminosities obtained with an extracted beam of 
$2 \times 10^5$ Pb/s for various target. (b) 
Yields per unit of rapidity  expected per LHC year with AFTER at mid rapidity 
with a 2.76 TeV Pb beam on various targets. Both are compared to the projected nominal luminosities and yield in Pb$p$ and PbPb runs
of the LHC at 8.8 and 5.5 TeV as well as in $d$Au and AuAu collisions at 200 GeV and 62 GeV at PHENIX~\cite{PHENIXdecadal} at RHIC. 
The yields are per LHC/RHIC year.  }\label{tab:yieldsPb}
\end{table}

The boost between the c.m.s. and the laboratory system is rather large, $\glabcms=\sqrt{s}/(2m_p)\simeq 38$, and the rapidity
shift is $\tanh^{-1} \blabcms\simeq 4.3$. The c.m.s. central-rapidity region --where the QGP is expected to be formed--, $y_{\rm c.m.s.}\simeq 0$, 
is thus highly boosted at an angle of 1.6 degrees with respect
to the beam axis  in the laboratory frame. The entire backward c.m.s. hemisphere ($y_{\rm .c.m.s}<0$) 
is easily accessible with standard experimental techniques. The forward hemisphere is less 
easily accessible because of the reduced distance from the (extracted) beam axis which  requires the use of highly segmented 
detectors to deal with the large particle density. One should be able to  access the region $-4.8\leq y_{\rm c.m.s.}\leq 1$ without specific difficulty.
 The main part of the particle yields could then be detected as well as high precision measurements in the whole backward hemisphere, down to 
the target rapidity. In particular, this should allow for a survey of observables at $y\simeq y_{\rm Pb\ target}$ in order to deepen
our understanding of the extended longitudinal scaling observed by Phobos~\cite{Back:2004je}.

In addition, let us emphasise that the  luminosities in the \pp\ and \pA\ mode~\cite{Brodsky:2012vg,Lansberg:2012wj} surpass 
those of RHIC by more than 3~orders of magnitude  
and are comparable to those of the LHC in the collider mode. 
In \pA, the nuclear target-species versatility provides a unique opportunity to study 
cold nuclear matter versus the features of the hot and dense matter formed in heavy-ion 
collisions, including the formation of the quark-gluon plasma.

\section{One example of physics studies: heavy quarkonia}

Even when these luminosities are translated in terms of yields for hard probes, AFTER is still competitive compared
to the LHC in the collider mode despite the significantly lower c.m.s energy.
\ct{tab:yieldsPb}(b) displays the expected \jpsi\ and $\Upsilon$ yields using 
the 2.76 TeV Pb beam on various targets. They are compared to those expected nominally\footnote{In both cases,
the numbers hold for one unit of rapidity with the branching into di-lepton but without
any acceptance nor efficiency corrections and without any suppression due to nuclear effects.} per year at 
RHIC in $d$Au and AuAu, at the LHC in Pb$p$  
and  in PbPb. 
As regards the $AA$ collisions, one sees that the yields in PbPb at $\sqrtS{NN}=72$ GeV are about equal to 
 those expected in a year at RHIC for AuAu at $\sqrtS{NN}=200$ GeV. They are  100 times larger than
those expected at 62 GeV and also similar (20\% lower, to be precise)  to that to be obtained 
during one LHC PbPb run at 5.5 TeV. The same global picture also applies for 
other quarkonium states --as well as for most of the hard probes for QGP studies, in particular 
open-charm and open-beauty.

AFTER is even more competitive for hard probe yields in \pp\ and \pA\ (see \cite{Brodsky:2012vg,Lansberg:2012kf}).
This provides a quarkonium and heavy-flavour observatory in \pp\ and \pA\ collisions where, by instrumenting 
the target-rapidity region, gluon and heavy-quark distributions of the proton, the neutron and the nuclei can 
be accessed at large $x$ -- even larger than unity in the nuclear case. For the first time, the far negative
$x_F$ region can be accessed where novel effects, such as those observed at large positive $x_F$ (see \eg~\cite{Hoyer:1990us}),
may be uncovered.

With the advent of modern detection technologies --such as those developed for instance for the ILC~\cite{Thomson:2009rp}--, 
one can be very hopeful that the study of all the quarkonium excited states is at reach. This is 
particularly true for the $\chi_c$ and $\chi_b$ resonances, even in the challenging high-multiplicity environment of \pA\ and \PbA\ 
collisions. Potential synergies with the project {\sc ChiC} \cite{EOI}
could therefore be very fruitful. The recent results on excited quarkonium states obtained 
at the LHC~\cite{Chatrchyan:2011pe,roberta-QM}  are particularly motivating for such studies at lower
energies.

High-statistics  data from \PbA,  \pA\ and \pp\ will be of great help to improve our understanding of heavy-quark and 
quarkonium production~\cite{Brambilla:2010cs}, to unravel cold from hot nuclear effects~\cite{ConesadelValle:2011fw} 
and to restore the status of heavy quarkonia as 
a golden-plated test~\cite{Matsui:1986dk} of lattice QCD in terms of dissociation temperature predictions at a \sqrtsNN\ where the 
procees of heavy-quark recombination  is expected to have a small impact. 

\section{Summary}

The physics reach of the most energetic heavy-ion accelerator ever built, the CERN-LHC, can be
significantly augmented by extracting a small fraction\ 
of its content --which would be lost in the collimators anyway-- with a bent crystal and by colliding it on a fixed target. The resulting energy in the c.m.s. 
amounts to $\sqrtS{NN}=72$ GeV in 
Pb$A$ collisions and $\sqrtS{NN}=115$ GeV in $pA$ collisions. Hydrogen and deuterium targets
also allow for $pp$ and $pd$ collisions to be studied with an extreme accuracy.
As expected for a fixed-target experiment, the resulting
luminosities surpass those of the existing relativistic heavy-ion colliders, \ie~RHIC and the LHC itself.

Such a fixed-target experiment, AFTER, would be a key actor in the study of the formation of the
quark-gluon plasma under scrutiny with  these heavy ion collisions.
Among the large number of proposed observables,  probes such as 
the suppression of quarkonia, the quenching of jets or the production of direct photons could easily be  accessed. 
Finally, let us stress that a fixed-target set-up also offers the novel opportunity of studying the QGP formation 
from the viewpoint of
one of the colliding nuclei.

\end{document}